\documentclass[conference]{IEEEtran}
\IEEEoverridecommandlockouts
\usepackage{cite}
\usepackage{algorithm}
\usepackage{algorithmicx} 
\usepackage[noend]{algcompatible}
\usepackage{amsmath,bm}
\usepackage[pdftex]{graphicx}
\usepackage{multirow}
\usepackage{array}
\usepackage[justification=centering]{caption}
\usepackage{epstopdf}
\usepackage{comment}
\usepackage[table,xcdraw]{xcolor}
\usepackage{booktabs} 
\usepackage{subfigure}
\usepackage{graphicx}
\newcommand\norm[1]{\left\lVert#1\right\rVert}
\newcommand*{\affaddr}[1]{#1}
\newcommand*{\affmark}[1][*]{\textsuperscript{#1}}

\def\BibTeX{{\rm B\kern-.05em{\sc i\kern-.025em b}\kern-.08em
    T\kern-.1667em\lower.7ex\hbox{E}\kern-.125emX}}
\def\baselinestretch{0.96}
\makeatletter
\newcommand\inlineeqno{\stepcounter{equation}\ (\theequation)}
\newcommand{\mathleft}{\@fleqntrue\@mathmargin0pt}
\newcommand{\mathcenter}{\@fleqnfalse}
\makeatother
\begin{document}

\title{DistPrivacy: Privacy-Aware Distributed Deep Neural Networks in IoT surveillance systems}
\author{%
	Emna Baccour\affmark[1], Aiman Erbad\affmark[1], Amr Mohamed\affmark[2],  Mounir Hamdi\affmark[1] and Mohsen Guizani\affmark[2].\\
	\affaddr{\affmark[1]College of Science and Engineering, Hamad Bin Khalifa University, Doha, Qatar.}\\
	\affaddr{\affmark[2]CS department, College of Engineering, Qatar University.}\\ 
}
\maketitle
\begin{abstract}
With the emergence of smart cities, Internet of Things (IoT) devices as well as deep learning technologies have witnessed an increasing adoption. To support the requirements of such paradigm in terms of memory and computation, joint and real-time deep co-inference framework with IoT synergy was introduced. However, the distribution  of Deep Neural Networks (DNN) has drawn attention to the privacy protection of sensitive data. In this context, various threats have been presented, including black-box attacks, where a malicious participant can accurately recover an arbitrary input fed into his device. In this paper, we introduce a methodology aiming to secure the sensitive data through re-thinking the distribution strategy, without adding any computation overhead. First, we examine the characteristics of the model structure that make it susceptible to privacy threats. We found that the more we divide the model feature maps into a high number of devices, the better we hide proprieties of the original image. We formulate such a methodology, namely DistPrivacy, as an optimization problem, where we establish a trade-off between the latency of co-inference, the privacy level of the data, and the limited-resources of IoT participants. Due to the NP-hardness of the problem, we introduce an online heuristic that supports heterogeneous IoT devices as well as multiple DNNs and datasets, making the pervasive system a general-purpose platform for privacy-aware and low decision-latency applications.
\end{abstract}


\section{Introduction}
Deep Neural Networks have become widely ubiquitous due to their ability to revolutionize a large variety of applications in many research fields, e.g., image recognition. Such high performance of DNN systems is related to the complex structure, the number of layers, and the significant resources consumed in the training and testing phases. More precisely, a typical DNN consists of tens of layers associated with thousands of neurons, incurring a computation and a memory occupation of a terabyte of floating point operations per second (flops) \cite{flop}. For example, VGG, which represents a state of the art performance on visual recognition, has 15 million neurons, 144 million parameters and 3.4 billion connections \cite{VGG}.

Because of their high requirements in terms of memory and  computation, DNN tasks have been mainly restricted to  highly  powerful machines,  e.g.,  cloud/edge servers. Meanwhile, IoT devices (e.g., sensors) were only responsible for collecting the data. Authors, in \cite{remote1}\cite{remote2}, decoupled the DNN network in a way that the computation is distributed between edge servers and conventional cloud. With such a server-centric approach, the large data volume (e.g., videos) and growing transmission latency have become problematic, particularly for decisions that require prompt intervention. Moreover,  system stability might be highly related to bandwidth availability. To address this challenge, deep learning tasks should be deployed as close as possible to the devices collecting the data. 

Nowadays, a pervasive environment is composed of thousands of computing devices distributed in different locations and able to connect together and form an IoT system. Some of these devices mostly execute light-weight tasks. This nexus between IoT units allows to distribute and push the DNN computation in close proximity of data sources, thus, reducing  bandwidth bottlenecks and incurring less latency and cost. 
More specifically, the deep neural network is divided into segments and each segment is allocated within a helper. Each helper shares the output to the next participant until generating the final prediction. The collaborative inference has gained the attention of the academia, particularly in mobile-cloud scenarios \cite{dis1,dis3,dis41}. In fact, to minimize the transmission overhead, many efforts have proposed to split the DNN network into two parts, where the first few layers are hosted in the IoT device while the rest are allocated in the remote servers to benefit from their high computation capacities. Still, bottleneck scenarios can occur. Recently, researchers have investigated the feasibility of using the limited resources of IoT devices to jointly allocate different segments of deep neural networks. In this way, the entire or most of the inference can be done at the proximity of data source \cite{dis6,dis8}. These efforts mainly explored the optimal partitioning of DNN that minimizes the shared data between devices. However, scheduling the participants to conduct complete inference tasks, while being constrained by computation and memory, was not taken into consideration by previous works.  Therefore, in order to support deep neural processing in IoT systems, the design of DNN distribution must be completely rethought by taking into account hardware and physical constraints, which will be done in our work.

\begin{figure}[h]
\centering
	\vspace{-0.2cm}
	\includegraphics[scale=0.52]{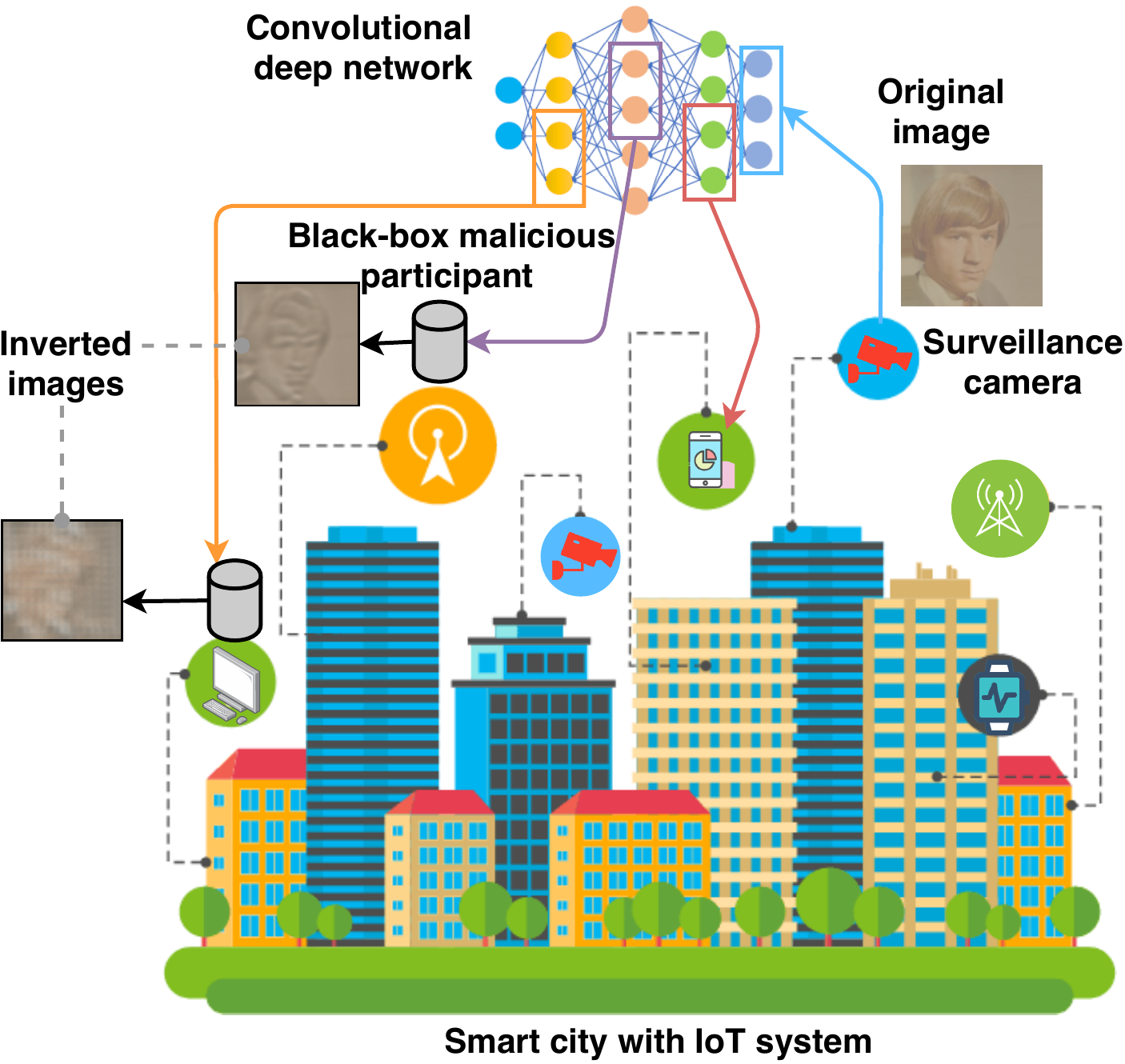}
	\caption{\small Illustration of the privacy-aware distributed CNN.}
	\label{model}
	\vspace{-0.35cm}
\end{figure}

Furthermore, the advancement and distribution of deep learning technologies have pointed out the security issues of sensitive data. Indeed, when the trained model is split and distributed among different helpers, an untrusted device can recover the input fed by the previous helper, even if it does not have insights about the model. Authors in \cite{security} proposed different attacks against collaborative inference, including black-box attacks. In black-box settings, the malicious participant only has  knowledge about his segment and attempts to design an inverse DNN network to map the received features to the targeted input to recover the original data. This work proved that attacking the inference is possible, when the neural system is distributed into layers as done in \cite{dis5}. Many countermeasures have been designed to enhance the privacy of deep networks. One of these privacy measures is adding noise to the intermediate outputs. However, if noise is added, the accuracy of the system will drop. Another way to secure the data is to perform inference on encrypted data. An obvious drawback of such approach is that it requires more computation, suffers from learning inefficiency, and cannot be applicable to all DNNs nor supported by all IoT devices. Finally, authors in \cite{security} proposed to divide the model after deep layers as the inversion becomes difficult. Knowing that IoT units have limited resources, this method can only be applied for larger capacity devices. In our work, we test black-box attacks on different distribution scenarios and we prove that distributing the neural network into layers is highly  fragile against inversion threats. Meanwhile, allocating parts of resultant feature maps into IoT devices contributes to hiding proprieties of the original image from untrusted participants. In this way, the black-box attack will be inefficient and unable to recover the private data, using only a small part of features, as illustrated in Figure \ref{model}. In this paper, we focus on surveillance IoT systems and we adopt Convolutional Neural Networks (CNNs), as an efficient DNN for image classification. As per our knowledge, we believe that we are the first to use CNN distribution to enhance the privacy of the system, without reducing the accuracy of results or increasing the computational load. 

The contribution of our study is three-fold: (1) First, we conduct deep empirical experiments to test the efficiency of black-box attack on different partition scenarios. In our simulation, we tested 4 different datasets and 4 state-of-the-art CNN networks. (2) We formulate our privacy-aware joint inference, namely DistPrivacy, as an optimization problem, aiming to minimize the classification latency, while considering the limited resources of participants and respecting the required privacy of the original data. (3) Finally, we propose an online heuristic that  supports  heterogeneous IoT  devices  as  well  as  multiple  DNNs, making the pervasive  system  a  general  platform  for  privacy-aware and  low latency  applications. Our  paper  is  organized  as  follows: Section \ref{deep_privacy} presents our proposed DistPrivacy framework, the empirical study, the problem formulation  and  the  online  heuristic.  The experimental  evaluation  is  provided  in  section  \ref{evaluation},  using  different state-of-the-art CNN networks. Finally, in section \ref{conclusion}, we draw the conclusions.
\section{Privacy-aware Distributed CNN for IoT devices}\label{deep_privacy}
\subsection{Black-box Inversion Attack Against distributed CNNs}
In this paper, we will handle the CNN privacy against black-box attacks. This type of attack assumes that the adversary has no knowledge about the weights or structure of the network. More specifically, without knowing the model parameters, the malicious participant cannot apply a gradient descent optimization to map the received input to the original image collected by the source device. The work in \cite{security2} proved, however, that the model can be reconstructed based only on information revealed from the received input. Even without collecting any insight, an inverse network can be trained to identify the inversion mapping from the input to the original image \cite{security}. Let $f^i_{\theta}(x)$ present a segment of the network serving to execute a part of the inference, where $x$ is the input image. Conceptually, the inverse network can be considered as $g=(f^{i}_{\theta})^{-1}$, trained with $in=f^i_{\theta}(x)$ as an input received from a previous device and $x$ as the inversion output. The loss function is expressed as: $\small{L(g^{(t)})=\frac{1}{k}\sum_{i=1}^k \norm{g(f_{\theta}(x_i))-x_i}^2_2}\inlineeqno$

Different from \cite{security} that recovered the output of full layers, we trained various networks to recover the original data, when having only few feature maps. The black-box attack is performed in three steps: (1) collect the training set of the inverse network by participating in the inference process: $(f^i_{\theta}(x_1),f^i_{\theta}(x_2),...,f^i_{\theta}(x_k))$; (2) train the network; and (3) recover the original image $x$ by inputting $in$ received from a previous participant. 
\begin{figure}[h]
	\vspace{-0.3cm}
	\mbox{
		\hspace{-0.4cm}
	     \subfigure[\label{black-box1}]{\includegraphics[scale=0.47]{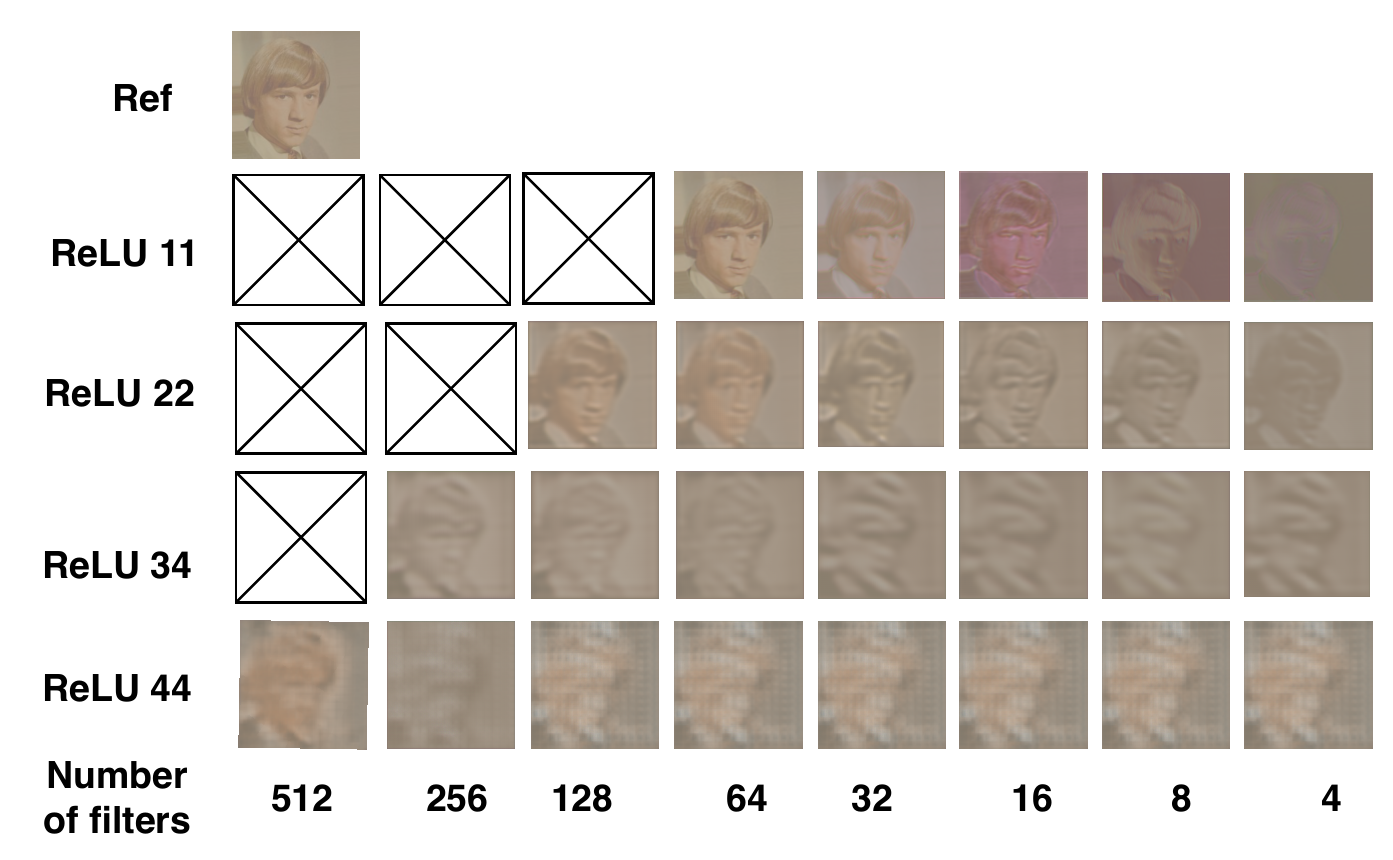}}
	     \hspace{-0.3cm}
		\subfigure[\label{black-box2}]{\includegraphics[scale=0.47]{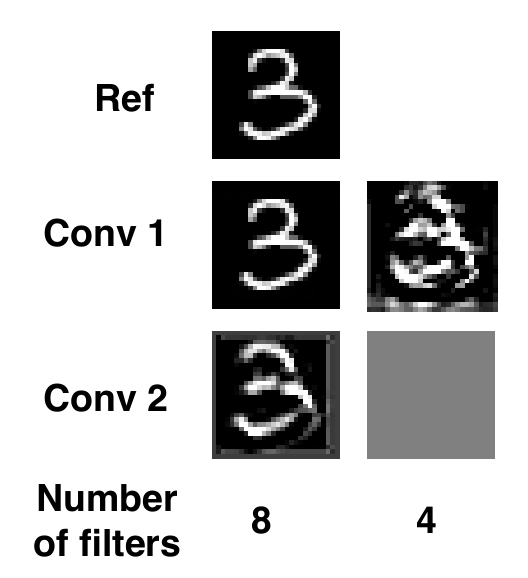}}
}\\\vspace{-0.2cm}
	\caption{\small Recovered images: (a) CELEBA, (b) MNIST.}
	\vspace{-0.2cm}
	\label{black-box}
\end{figure}

We applied the black-box attack on four standard CNN benchmark datasets: CIFAR10 trained with a small CNN (6 convolutional and 2 fully connected layers), MNIST trained with LeNet (2 convolutional and 3 fully connected layers), CELEBA with VGG19 (16 convolutional  and 3 fully connected layers), and Stanford CAR with VGG16 (13 convolutional and 3 fully connected layers). Without loss of generality, in Figure \ref{black-box}, we present the inversion results of MNIST and CELEBA, for layers where the number of filters changes (e.g., conv22/ReLU22 receives 64 feature maps and outputs 128 feature maps). Accordingly, Table \ref{ssim} shows the average Structural Similarity Index (SSIM) between original and recovered images from different datasets and under different configurations, when a participant owns only few filters. The SSIM is a metric that measures the image quality degradation and the similarity compared to the original sample.
\begin{table}[]
\center
\footnotesize
\tabcolsep=0.11cm
\setlength\extrarowheight{-2pt}
\caption{\small SSIM of black-box inversion against different datasets.}
\label{ssim}
\begin{tabular}{|l|l|l|l|l|l|l|l|l|l|}
\hline
\begin{tabular}[c]{@{}l@{}}Nb filters\\ per device/\\Layer\end{tabular} & 512 & 256 & 128 & 64 & 32 & 16 & 8 & 4 & 2 \\ \hline
\begin{tabular}[c]{@{}l@{}}CIFAR\\ ReLU 11\end{tabular} & \cellcolor[HTML]{C0C0C0}{\color[HTML]{EFEFEF} } & \cellcolor[HTML]{C0C0C0}{\color[HTML]{EFEFEF} } & \cellcolor[HTML]{C0C0C0}{\color[HTML]{EFEFEF} } & 0.99 & 0.6 & 0.56 & 0.4 & 0.3 & 0.26 \\ \hline
\begin{tabular}[c]{@{}l@{}}CIFAR\\ ReLU22\end{tabular} & \cellcolor[HTML]{C0C0C0} & \cellcolor[HTML]{C0C0C0} & 0.86 & 0.7 & 0.49 & 0.34 & 0.13 & 0.1 & 0.07 \\ \hline
\begin{tabular}[c]{@{}l@{}}CIFAR\\ ReLU32\end{tabular} & \cellcolor[HTML]{C0C0C0}{\color[HTML]{EFEFEF} } & \cellcolor[HTML]{C0C0C0}{\color[HTML]{EFEFEF} } & 0.6 & 0.51 & 0.41 & 0.18 & 0.08 & 0.07 & 0.01 \\ \hline
\begin{tabular}[c]{@{}l@{}}MNIST\\ Conv1\end{tabular} & \cellcolor[HTML]{C0C0C0}{\color[HTML]{EFEFEF} } & \cellcolor[HTML]{C0C0C0}{\color[HTML]{EFEFEF} } & \cellcolor[HTML]{C0C0C0}{\color[HTML]{EFEFEF} } & \cellcolor[HTML]{C0C0C0}{\color[HTML]{EFEFEF} } & \cellcolor[HTML]{C0C0C0}{\color[HTML]{EFEFEF} } & \cellcolor[HTML]{C0C0C0}{\color[HTML]{EFEFEF} } & 0.99 & 0.28 & \cellcolor[HTML]{C0C0C0}{\color[HTML]{EFEFEF} } \\ \hline
\begin{tabular}[c]{@{}l@{}}MNIST\\ Conv2\end{tabular} & \cellcolor[HTML]{C0C0C0}{\color[HTML]{EFEFEF} } & \cellcolor[HTML]{C0C0C0}{\color[HTML]{EFEFEF} } & \cellcolor[HTML]{C0C0C0}{\color[HTML]{EFEFEF} } & \cellcolor[HTML]{C0C0C0}{\color[HTML]{EFEFEF} } & \cellcolor[HTML]{C0C0C0}{\color[HTML]{EFEFEF} } & \cellcolor[HTML]{C0C0C0}{\color[HTML]{EFEFEF} } & 0.73 & 0 & \cellcolor[HTML]{C0C0C0}{\color[HTML]{FFFFFF} } \\ \hline
\begin{tabular}[c]{@{}l@{}}CELEBA\\ ReLU11\end{tabular} & \cellcolor[HTML]{C0C0C0}{\color[HTML]{EFEFEF} } & \cellcolor[HTML]{C0C0C0}{\color[HTML]{EFEFEF} } & \cellcolor[HTML]{C0C0C0}{\color[HTML]{EFEFEF} } & 0.96 & 0.81 & 0.66 & 0.27 & 0.09 & 0.1 \\ \hline
\begin{tabular}[c]{@{}l@{}}CELEBA\\ ReLU22\end{tabular} & \cellcolor[HTML]{C0C0C0}{\color[HTML]{EFEFEF} } & \cellcolor[HTML]{C0C0C0}{\color[HTML]{EFEFEF} } & 0.76 & 0.69 & 0.71 & 0.59 & 0.59 & 0.4 & 0.4 \\ \hline
\begin{tabular}[c]{@{}l@{}}CELEBA\\ ReLU34\end{tabular} & \cellcolor[HTML]{C0C0C0}{\color[HTML]{EFEFEF} } & 0.56 & 0.51 & 0.47 & 0.49 & 0.46 & 0.45 & 0.45 & 0.45 \\ \hline
\begin{tabular}[c]{@{}l@{}}CELEBA\\ ReLU44\end{tabular} & 0.26 & 0.39 & 0.3 & 0.3 & 0.3 & 0.3 & 0.3 & 0.3 & 0.3 \\ \hline
\begin{tabular}[c]{@{}l@{}}CAR\\ ReLU11\end{tabular} & \cellcolor[HTML]{C0C0C0}{\color[HTML]{EFEFEF} } & \cellcolor[HTML]{C0C0C0}{\color[HTML]{EFEFEF} } & \cellcolor[HTML]{C0C0C0}{\color[HTML]{EFEFEF} } & 0.98 & 0.92 & 0.93 & 0.88 & 0.69 & 0.04 \\ \hline
\begin{tabular}[c]{@{}l@{}}CAR\\ ReLU22\end{tabular} & \cellcolor[HTML]{C0C0C0}{\color[HTML]{EFEFEF} } & \cellcolor[HTML]{C0C0C0}{\color[HTML]{EFEFEF} } & 0.83 & 0.74 & 0.59 & 0.47 & 0.5 & 0.4 & 0.26 \\ \hline
\begin{tabular}[c]{@{}l@{}}CAR\\ ReLU33\end{tabular} & \cellcolor[HTML]{C0C0C0}{\color[HTML]{EFEFEF} } & 0.68 & 0.58 & 0.58 & 0.55 & 0.46 & 0.31 & 0.18 & 0.18 \\ \hline
\begin{tabular}[c]{@{}l@{}}CAR\\ ReLU43\end{tabular} & 0.36 & 0.33 & 0.30 & 0.36 & 0.36 & 0.31 & 0.29 & 0.34 & 0.33 \\ \hline
\end{tabular}
\vspace{- 6 mm}
\end{table}

Several conclusions can be drawn based on the illustrative Figure \ref{black-box} and Table \ref{ssim}. First, the malicious participant can recover the input image using black-box attack on the shallow layers, if he/she receives all feature maps generated from the previous layer. For example, the recovered images at ReLU11 or ReLU22 of CELEBA  have a high quality. When attacking the network at deep layers, the inverted sample becomes vague, for all datasets. To secure the network from inversion attack, we propose that a participant does not receive the whole output produced by the previous device, as done in \cite{dis5}. Instead, the output feature maps will be distributed to more than one IoT participant. Then, these untrusted participants apply the attributed tasks (conv, ReLU, etc) on the received segments, without being able to fully recover the original image. As illustrated in Table \ref{ssim}, distributing the output of the shallow layers into multiple devices contributes to reducing the possibility to recover the images. Particularly, when sharing the output of the layer ReLU11/CELEBA with two devices (32 maps each) or four devices (16 each), the original image becomes noisy, but still can be distinguished (Figure \ref{black-box1}). When distributing the output into 8 participants (8 maps each), recovering the image becomes not feasible. As we go deeper in the network, retrieving the original data becomes more difficult and a lower number of IoT devices is needed to distribute the output and maintain the system security. Therefore, in the rest of the paper, we will establish a trade-off between the privacy level and the shared data load between different IoT participants to execute the inference, while minimizing the latency. Without loss of generality, the level of privacy will be calculated in terms of SSIM. Note that the metric to evaluate image degradation could vary depending on the dataset. 
\subsection{System Model}
Our surveillance system comprises a set of image-generating IoT units (e.g., Cameras), namely $\mathcal{S}=\{s_1,...,s_t\}$, capturing images to be classified by $N$ CNNs. Each IoT device can request to run only one type of inference (e.g., Face recognition) and each image has $ch$ channels ($ch=3$ for RGB images). Let $\mathcal{I}=\{I_1,...,I_D\}$ denote the set of $D$ heterogeneous IoT devices used to compute different segments of the convolutional neural network. Without loss of generality, we assume that the image-generating devices do not participate in computing intermediate segments of the CNN. Different IoT participants $I_i$ are characterized by limited resources, including memory usage $\Bar{m_i}$, computation capacity $\Bar{c_i}$, and bandwidth availability $\Bar{b_i}$.

In this paper, we assume that all devices communicate through the same transmission technology (eg., Wi-Fi or cellular) with different transmission rates $\rho_i$. We note that the study of the wireless channel impact on the data is out of the scope of this work. Finally, we assume that all devices are not isolated and they can all reach each others. Else, a constraint on the communication range can be added.

As described previously, $N$ CNNs are deployed in our IoT system, one network for each source $s_i$. Let $L_{n_j}$  denote the number of layers of the CNN $n_j \in \{1....N\}$. Each layer $l^k_{n_j} \in \{1,...,L_{n_j}\}$ results in $P_{l^k_{n_j}}$ feature maps. The output of the layer $l^k_{n_j}$ will be considered as the number of segments to be computed by the next layer $l^{k+1}_{n_j}$. Each segment is characterized by a computation requirement $c^{k,p}_j$ and a memory demand $m^{k,p}_j$.  Without loss of generality, the computational load is measured as the number of multiplications required to accomplish the layer goals. Accordingly, the computation of layers that do not run any multiplication (e.g., ReLU, maxpool) will be neglected \cite{b1}. The computation requirement of a convolution layer segment $p_{l^k_{n_j}}$ can be calculated as follows:
$\label{eq:1} \qquad \qquad \qquad c^{k,p}_j=S^2_{k+1}.P_{l^{k+1}_{n_j}}.o_{k+1}^2, \inlineeqno$
\newline
where $S_{k+1}$ denotes the spatial size of the filter corresponding to the layer $l^{k+1}_{n_j}$ and $o_{k+1}$ represents the spatial size of the output feature map.
The computation requirement of a fully-connected layer can be calculated as follows: $ \label{eq:2} \qquad \qquad \qquad c^k_j=n^*_{k-1}.n^*_k, \inlineeqno$
\newline
$n^*_k$ represents the number of neurons of the layer $l^{k}_{n_j}$ and $n^*_{k-1}$ is the number of neurons of the layer $l^{k-1}_{n_j}$. We note that, in our work, the fully connected layers will be computed on a single device as the output of such layers cannot be recovered \cite{security}. Accordingly, $\{P_{l^{k}_{n_j}},...,P_{l^{L}_{n_j}}\}$ are equal to 1 if  $l^{k}_{n_j}$ is a fully connected layer. The memory demand of $p_{l^{k}_{n_j}}$ can be calculated as the number of stored weights multiplied by the memory word-length (number of bits required to store each weight) \cite{b1}: $\qquad \quad  \label{eq:3}  m^{k,p}_j=W^{k,p}_j.b, \inlineeqno$
\newline
Typically, $b$ is equal to 4 bits if the single-precision flop data type is used \cite{b3}. Finally, let $O^k_{i1,i2}$ denote the memory occupation of the output data communicated between the participant $I_{i1}$ and $I_{i2}$ to offload segments of the layer  $l^k_{n_j}$.

Next, we will introduce the optimal placement of different segments in the IoT units participating in our system, while taking into consideration the privacy against  black-box attacks. The optimization is executed periodically to cover the variation of the network, and the inclusion and removal of participants.
\subsection{Problem formulation}
\vspace{-0.05cm}
The proposed privacy-aware strategy relies on one decision variable, namely $A^i_{r^*,l,p}$. $A^i_{r^*,l,p}$ is equal to 1, if the helper $I_i$ computes the feature map $p$ at the layer $l$ of the request $r$; 0 otherwise.
We note that $r \in RQ$ presents the request for inference computation. In practice, $r$ represents the index of the source device that requested the classification of an image. We remind that each device is related to a single CNN that we will denote by $r^*$. 

The objective function models the total latency of computing the set of requests $RQ$, in the distributed IoT system. This latency is defined as the time necessary to transfer the output of layers between participants and to compute different tasks, and it is expressed as follows:
\begin{equation}\label{obj}
\footnotesize
    \begin{aligned}
  L_{IoT}=\sum\limits_{r \in RQ}\sum\limits_{l=1}^{L_{r^*}} max( \frac{O^{l-1}_{i,j}}{\rho_i}+t^{r^*,l,j}_c, \forall I_i,I_j \in  \mathcal{I}\cup \mathcal{S}), 
    \end{aligned}
\end{equation}
\begin{equation}\label{O}
\footnotesize
    \begin{aligned}
\text{where: }O^{l}_{i,j}= [(C^l*o_{l}^2*min(1,\sum\limits_{p_1=1}^{P_{l-1}}A^{i}_{r^*,l,p_1})*\sum\limits_{p_2=1}^{P_{l}}A^{j}_{r^*,l+1,p_2}+\\(Ac^l*o_{l}^2+F^l*n^*_{l})*\sum\limits_{p_1=1}^{P_{l-1}}A^{i}_{r^*,l,p_1}A^{j}_{r^*,l+1,p_1})*b]*\mathrm{1}_{i\neq j}\quad 
    \end{aligned}
\end{equation}
The binary variable $C^l$ is equal to 1, if the layer $l$ is a conv layer; $Ac^l$ is equal to 1, if $l$ is an activation layer or a Maxpool; and $F^l$ is equal to 1, if $l$ is a fully connected layer. 
The objective function in Eq. (\ref{obj}) is composed of 2 parts:

(1) The transmission latency of intermediate segments $p$ between different IoT participants, which is presented by $\frac{O^{l-1}_{i,j}}{\rho_i}$: We remind that $\rho_i$ defines the data-rate of the considered transmission technology equipped with $I_i$ and $O^{l-1}_{i,j}$ is the size of $l-1$ segments shared between $I_i$ and $I_j$. $O^{l-1}_{i,j}$ is equal to 0, if $I_i = I_j$, including the case of a source device generating the image and processing the first layer.

(2) The processing latency of different segments of layer $l$ on the IoT participants, which is expressed as follows:
\begin{equation}\label{tc}
\footnotesize
    \begin{aligned}
t_c^{r^*,l,j}=\sum\limits_{p=1}^{P_{l-1}} A^j_{r^*,l,p} *\frac{c_{r^*}^{l,p}}{e(j)}, \qquad \forall \quad I_j \in \mathcal{I}\cup \mathcal{S}
    \end{aligned}
\end{equation}

The computation time of the feature map $p$ on the $I_j$ unit is approximated as the ratio between the computational demand $c_{r^*}^{l,p}$ required by the segment, and the number of multiplications $e(j)$ the unit $I_j$ is able to carry out in one second \cite{dis5}. In practice, $e(j)$ indirectly includes the available cores per device to parallelize operations.  Furthermore, the offloading of different output feature maps of the same layer to next participants and their computation are done synchronously. Thus, the transmission and processing time is defined as the larger latency among different participants to receive and accomplish the layer task. Ultimately, our privacy-aware distributed deep convolutional network can be formulated as follows:

\begin{subequations}
\footnotesize
	\label{optimal}
	\allowdisplaybreaks
		\begin{align}
	\allowdisplaybreaks
		\begin{split}
    \underset{ \begin{subarray}{c}
    (A^i_{r^*,l,p}) 
    \end{subarray}}\min \qquad L_{IoT} \qquad \qquad \qquad \qquad 
	\label{eq:optProb} 
	\end{split}\\
	\begin{split}
\text{s.t} \quad \sum\limits_{r \in RQ}\sum\limits_{l=1}^{L_{r^*}}\sum\limits_{p=1}^ {P_{l-1}}A^i_{r^*,l,p}*m^{l,p}_{r^*}\leq \bar{m_i} \quad \forall  I_i \in \mathcal{I}\cup\{s_1..s_n\},
	\label{eq:constraint1}
	\end{split}\\
		\begin{split}
\qquad \sum\limits_{r \in RQ}\sum\limits_{l=1}^{L_{r^*}}\sum\limits_{p=1}^ {P_{l-1}} A^i_{r^*,l,p}*c_{r^*}^{l,p}\leq \bar{c_i} \quad \forall  I_i \in \mathcal{I}\cup\{s_1..s_n\},
	\label{eq:constraint2}
	\end{split}\\
			\begin{split}
\qquad \sum\limits_{r \in RQ}\sum\limits_{l=1}^{L_{r^*}}\sum\limits_{p=1}^ {P_{l-1}} \sum\limits_{I_j \in \mathcal{I}} O^{l}_{i,j}\leq \bar{b_i} \quad \forall  I_i \in \mathcal{I}\cup\{s_1..s_n\}
	\label{eq:constraint21}
	\end{split}\\
			\begin{split}
\qquad \sum\limits_{i \in \mathcal{I}\cup\{s_1..s_n\}} A^i_{r^*,l,p}= \Big\{
    \begin{array}{ll}
        1 & \mbox{if } l \leq L_{c_{r^*}}, p \leq P_{l-1} \\
        0 & \mbox{Otherwise} 
    \end{array}
	\label{eq:constraint3}
	\end{split}\\
		\begin{split}
\qquad \sum\limits_{p=1}^ {P_{l-1}} A^i_{r^*,l,p}\leq Nf^l(SSIM)\quad \forall  I_i \in \mathcal{I},   l \leq SP_{r^*}(SSIM)
	\label{eq:constraint5}
	\end{split}\\
			\begin{split}
\qquad \sum\limits_{i \in \mathcal{I}\cup\{s_1..s_n\}}\Pi_{p=1}^ {P_{l-1}}A^{i}_{r^*,l,p} \geq F^l.\neg F^{l-1}, \quad \forall l \leq L_{c_{r^*}}
	\label{eq:constraint10}
		\end{split}\\	
			\begin{split}
\qquad A^{r}_{r^*,l,p}=\Bigg\{
    \begin{array}{ll}
        1 & \mbox{if } l \in \{1,L_{r^*}\} \\
        F^l.\neg F^{l-1} & \mbox{if } SP_{r^*}(SSIM) \leq l \\
        0 & \mbox{Otherwise}
    \end{array} \\
    \forall l \in \{1...L_{r^*}\}, p \in \{1..P_{l-1}\}
	\label{eq:constraint31}
		\end{split}\\
	\begin{split}
	\qquad A^i_{r,l,p} \in \{0,1\} 
	\label{eq:constraint9} 
	\end{split} 
	\end{align}
\end{subequations}
Equations (\ref{eq:constraint1}) and (\ref{eq:constraint2}) ensure that the constraints on computational load and memory usage are respected for each IoT participant or source device. Similarly,  Eq. (\ref{eq:constraint21}) verifies that the total transmitted output of the computed segments respects the available bandwidth. Next, equation (\ref{eq:constraint3}) guarantees that the computation of each segment $p$ is assigned to only one node and (\ref{eq:constraint5}) ensures that the number of feature maps $Nf^l$ received by any IoT device cannot allow the malicious participant to recover the data with an accuracy larger than the maximum tolerated SSIM. To illustrate the security constraint, we suppose that the tolerated SSIM is equal to 0.4. If we conduct an inference on CIFAR (Table \ref{ssim}),  ReLU11 and ReLU22 layers should be distributed on 8 devices ($Nf^{11}(SSIM)=8, Nf^{22}(SSIM)=16$), and ReLU32  on 4 units ($Nf^{32}(SSIM)=32$). We verify this constraint only for the layers preceding the split point $SP_{r^*}(SSIM)$, as  next layers are recovered with a lower SSIM than tolerated, even when receiving all filters.
Finally, Eq. (\ref{eq:constraint31}) ensures that the source devices do not participate in computing intermediate segments and only handle the first and the last layers. We emphasize that, the first and last layers are always computed in the source device to protect the privacy of the original image and the classification results. Also, as we chose not to distribute fully connected layers $F^l$, all segments from the non fully connected previous layer $\neg F^{l-1}$ will be offloaded to one device (constraint (\ref{eq:constraint10})), which exposes the data to inversion risks, if $l$ is lower than $SP_{r^*}$ (case of MNIST). Therefore, in such a scenario, constraint (\ref{eq:constraint31}) ensures that the first fully connected layer is computed on the source device. 
\subsection{DistPrivacy: Online heuristic}
The optimization in (\ref{optimal}) is an NP-hard problem, which makes finding the optimal solution extremely challenging in terms of time. Because  of  this combinatorial  complexity and the real-time requirements of the classification, we propose an online greedy algorithm, namely DistPrivacy, illustrated in algorithm \ref{DistPrivacy}. In fact, when receiving an inference request, different tasks of the CNN network are distributed among participants guided by the following rules: all segments of the first and the last layers are computed in the source device, along with the first fully connected layer, in case the tolerated SSIM is not reached at this level ($\text{line 9}  \rightarrow \text{line 12} $). Segments of intermediate layers are handled by un-trusted participants ($\text{line 13} \rightarrow \text{line 23}$).  Particularly, to align with the problem formulation, the helper that achieves lower latency should be selected. However, the chosen device can suffer from bandwidth shortage and consequently cannot share the resultant  feature maps with next participants. Therefore, since the greedy allocation does not have an overview of bandwidth requirements, the selected helper is the one that has the lowest $nrm(j)= \alpha.\tilde{t(j)}+\beta. \tilde{1/\bar{b_j}}$ (line 16). $nrm$ establishes a trade-off between the latency and the eventual available bandwidth. Without loss of generality, we fixed $\alpha$ and $\beta$ to $0.7$ and $0.3$, respectively. This participant should respect the memory and computation availability ($cond1$) and the privacy level constraint ($cond4$). Furthermore, the bandwidth within participants that computed the layer $l$-1, should be available to accomplish the transmission of segments to the current layer $l$ ($cond2$). Rejections will occur, if resources are exhausted.
\begin{algorithm}[h]
\caption{DistPrivacy}
\label{DistPrivacy}
\footnotesize
\begin{algorithmic}[1]
\State \textbf{Input:} $A^i=0$, $\bar{c_i}, \bar{m_i},\bar{b_i}, e(i), \rho_i, \forall I_i \in \mathcal{I}$, $Nf^l$,\quad$L_{IoT}=0$.
\State $cond1_j^{r,l,p}$=$c_{r^*}^{l,p}\leq \bar{c_j} \quad \& \quad m_{r^*}^{l,p}\leq \bar{m_j}$.
\State $cond2^{r,l-1,p}_j$=$(O^{l-1,p}_{i,j}\leq \bar{b_{i}} \quad \forall I_i \in \mathcal{I})$.
\State $cond3^{r,l}$=$l<SP_{r^*}(SSIM) \quad \& \quad F^l=1 \quad \& \quad F^{l-1}=0$
\State $cond4^{r,l,p}_j$=($\sum\limits_{s=1}^ {p-1} A^{j}_{r^*,l,s}\leq Nf^l(SSIM) \quad || \quad l> SP_{r^*}(SSIM))$
\FOR{each $r \in RQ$ }
\FOR{each $l \in \{1..L_{r^*}\}$ }
\FOR{each $p \in \{1..P_{l-1}\}$}
\IF {$(l=1 || (l=L_{r^*}\quad \& \quad cond2_r^{r,L_{r^*}-1,p}) || \quad\quad\quad(l>1\quad\&\quad cond3_r^{r,l,p})) \quad \& \quad cond1_r^{r,l,p} $}
\State $A^r_{r^*,l,p}=1$
\State $\bar{c_r}=\bar{c_r}-c_{r^*}^{l,p}, \bar{m_r}=\bar{m_r}-m_{r^*}^{l,p},\bar{b_{i}}=\bar{b_{i}}-O^{l-1,p}_{i,r}$
\ELSE
\State selected =0
\FOR{each $I_j \in \mathcal{I}$ }
\State t(j)=$max( \frac{O^{l-1,p}_{i,j}}{\rho_i}, \forall I_i \in  \mathcal{I})+ c_{j}^{l,p}/e(j)$
\State $nrm(j)= \alpha.\tilde{t(j)}+\beta. \tilde{1/\bar{b_j}}$
\ENDFOR
\WHILE{$nrm \neq \emptyset \quad \& selected =0$}
\State $[pt,nrm(pt)] = min(nrm)$
\IF{$cond1_{pt}^{r,l,p} \& \quad cond2_{pt}^{r,l-1,p} \& \quad cond4_{pt}^{r,l,p}$}
\State $A^{pt}_{r^*,l,p}=1$
\State $\bar{c_{pt}}=\bar{c_{pt}}-c_{r^*}^{l,p}, \bar{m_{pt}}=\bar{m_{pt}}-m_{r^*}^{l,p},\bar{b_{i}}=\bar{b_{i}}-O^{l-1,p}_{i,pt}$
\State $selected=1$
\ELSE{\quad Remove nrm(pt)}
\ENDIF
\ENDWHILE
\IF{$nrm = \emptyset$}
\State \quad rejected=rejected+1
\ENDIF
\ENDIF
\ENDFOR
\State $L_{IoT}=L_{IoT}+max( \frac{O^{l-1}_{i,j}}{\rho_i}+t^{r^*,l,j}_c, \forall I_i,I_j \in \mathcal{I} \cup \mathcal{S})$
\ENDFOR
\ENDFOR
\end{algorithmic}
\end{algorithm}
\vspace{-0.2cm}
\section{Performance evaluation}\label{evaluation}
In this section, our DistPrivacy system is evaluated under different networking capacities. Particularly, the impact of the number of IoT devices, their capacities, and the type of requests, on the total latency and shared data is studied. DistPrivacy has been validated on four  benchmark CNNs in a surveillance scenario of image classification to control a critical area. The source devices capture 36x36 RGB sized images (CIFAR), 28x28 gray images (MNIST) and 128x128 RGB images (CELEBA and Stanford CAR). Moreover, our pervasive IoT system consists of three technological families of devices, i.e, Raspberry Pi B+, LG Nexus 5 and STM32H7. The former device is equipped with 1.4GHz 64-bit quad-core processor and 1GB RAM, the second device is considered as a powerful unit having 2.28 GHz processor and 2GB RAM, and the last one, which is a small device (e.g,. smart watch), is endowed with 400 MHz-cortex and 1 MB RAM. The number of multiplications per second $e$,  defined as the tenth of the clock cycles per number of cores \cite{dis5}, is equal to 560 for RPi3, 800 for LG Nexus and 40 for STM32H7. The small IoT devices are equipped with a low bandwidth technology (IEEE 802.11ah), having a data rate $\rho$ equal to 7.2 Mb/s. Meanwhile, the powerful machines (RPi3 and LG Nexus) are endowed with IEEE 802.11n standard with a   $\rho$ equal to 72.2 Mb/s. We note that all cameras are deployed with RPi3 system. 

The proposed system is first evaluated on the data-collection to decision-taking latency, for different privacy levels. The measured latency is defined as the time between acquiring the image and obtaining the classification, while the privacy level is presented by the SSIM metric (see table \ref{ssim}). Second, we evaluate the data shared between all participants to accomplish all inferences. This data load includes indirectly the cost of the network. In our simulation, we generate 320 requests through a Poisson process with a data rate $\lambda$ equal to 3.

\begin{figure}[h]
	\mbox{
	\hspace{-0.4cm}
	     \subfigure[\label{nbl}]{\includegraphics[scale=0.47]{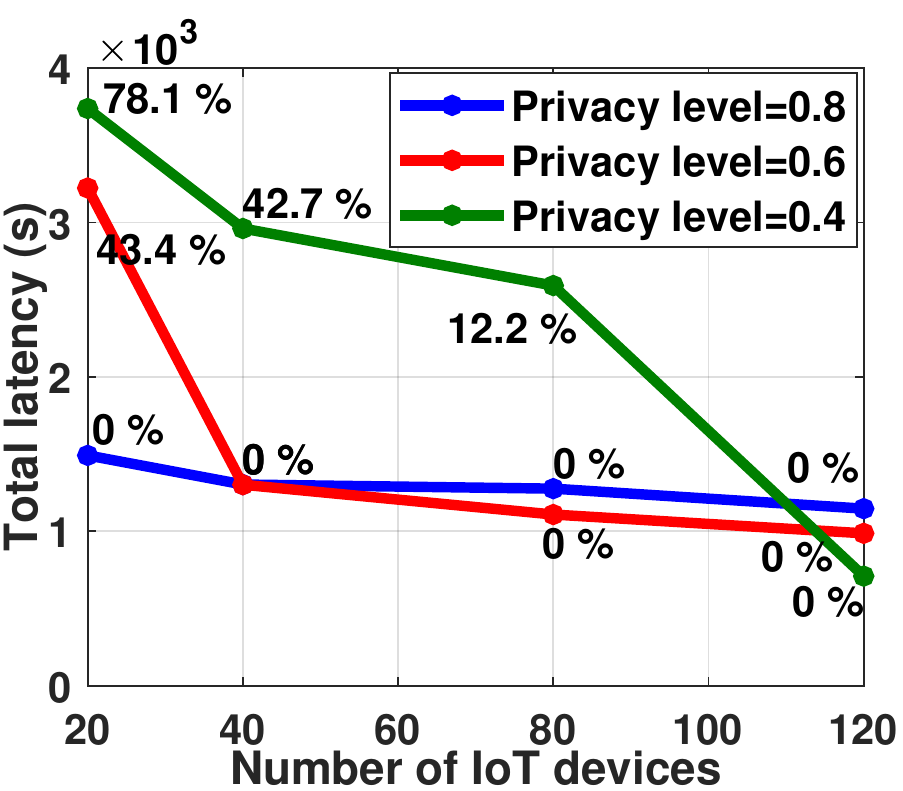}}
	     \hspace{-0.3cm}
	     \subfigure[\label{nbd}]{\includegraphics[scale=0.466]{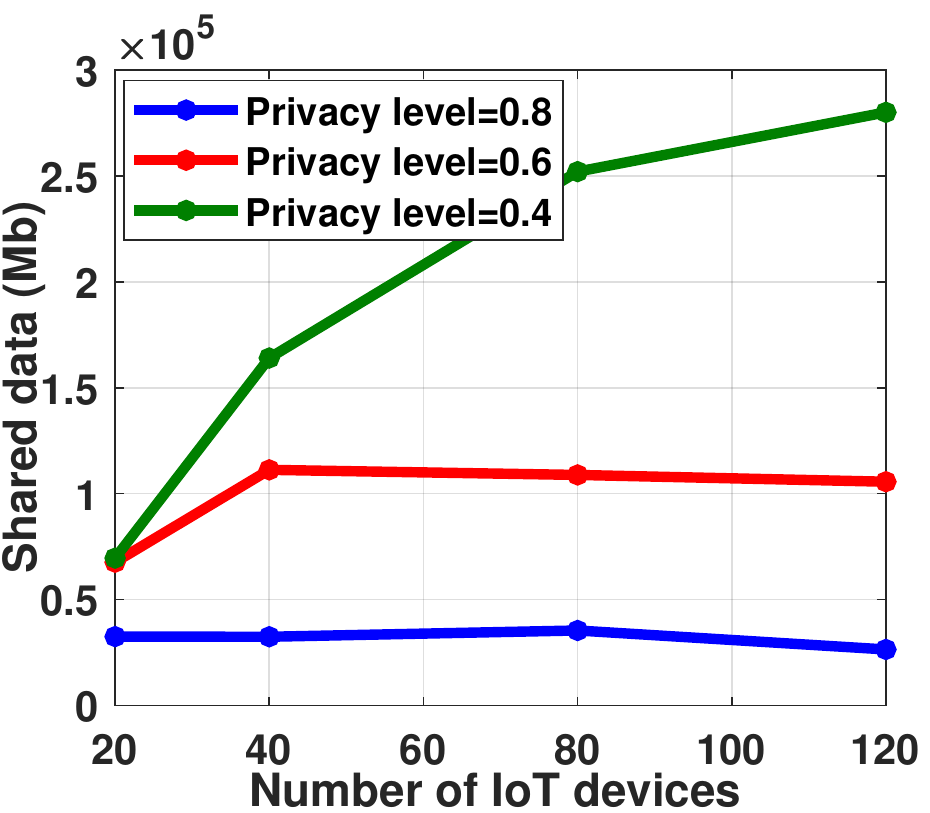}}
}\vspace{-0.25cm}\\
	\mbox{
	\hspace{-0.4cm}
	    \subfigure[\label{disl}]{\includegraphics[scale=0.45]{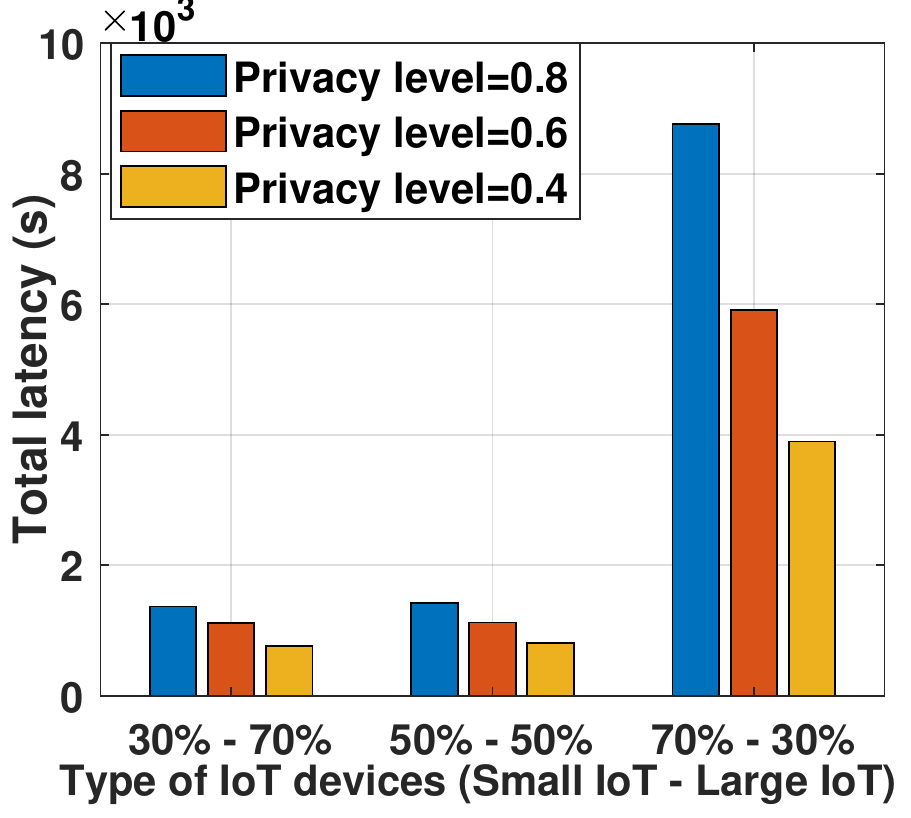}}
	    \hspace{-0.3cm}
        \subfigure[\label{disd}]{\includegraphics[scale=0.46]{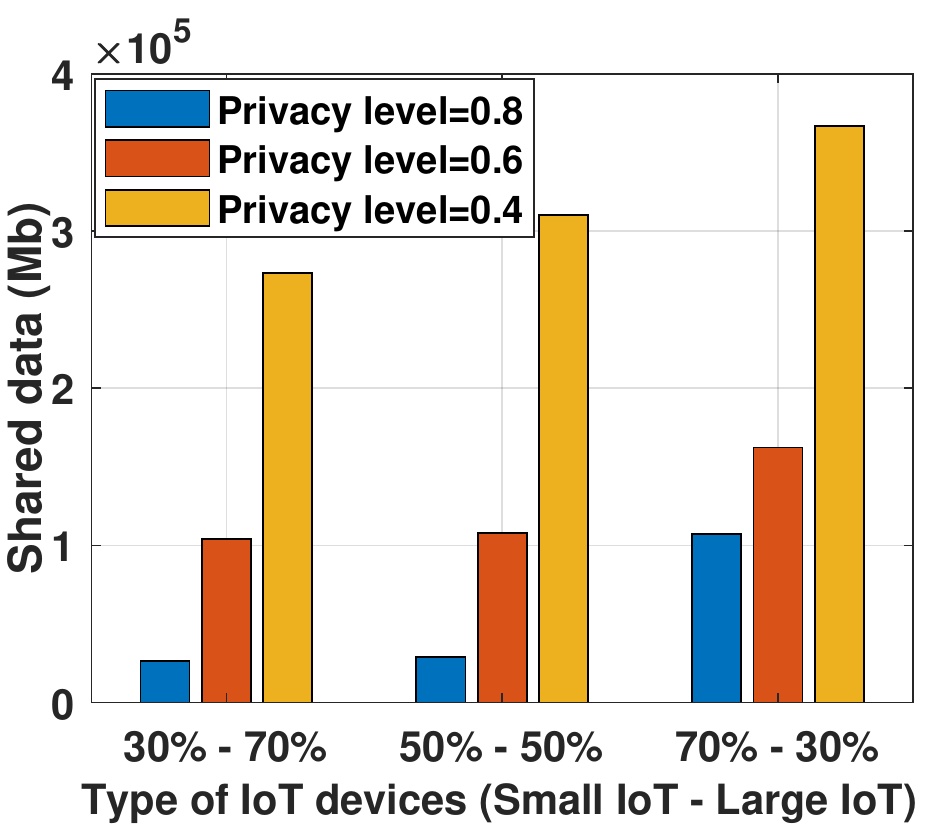}}
}\vspace{-0.2cm}\\

	\mbox{
	\hspace{-0.5cm}
		\subfigure[\label{typel}]{\includegraphics[scale=0.45]{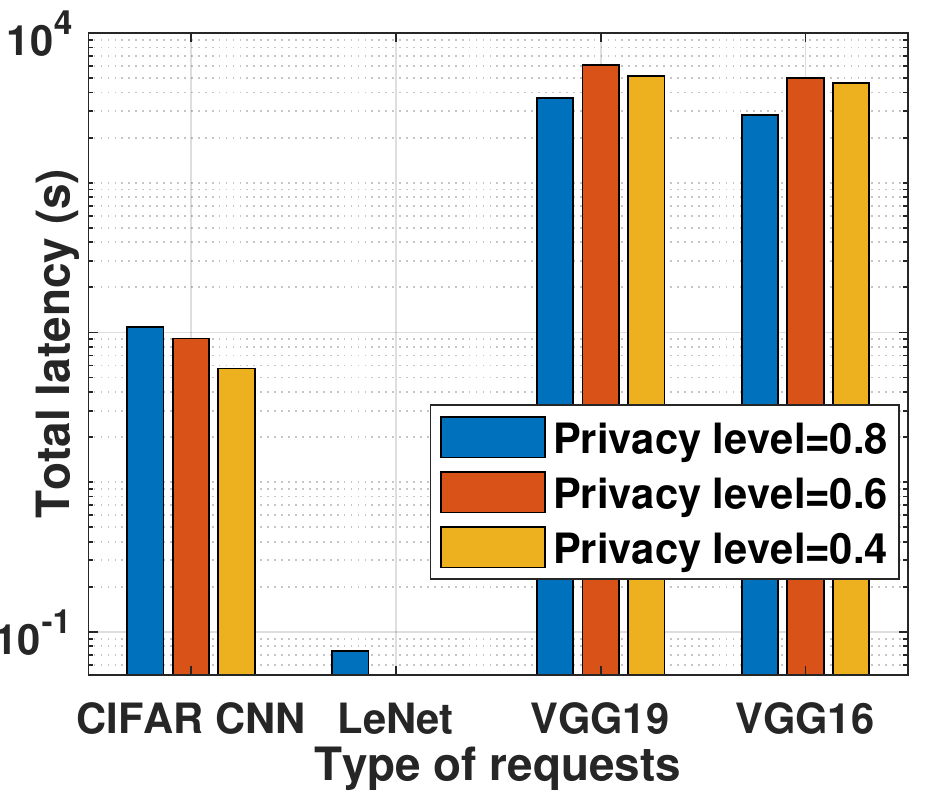}}
		\hspace{-0.2cm}
		\subfigure[\label{typed}]{\includegraphics[scale=0.47]{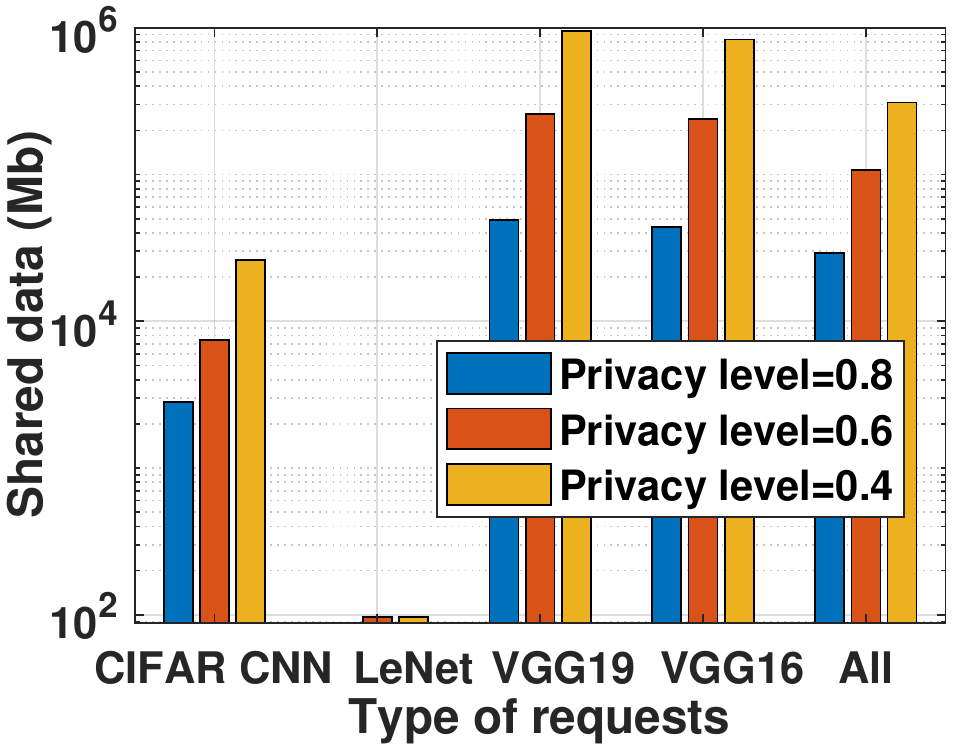}}
}
\vspace{-0.2cm}
	\caption{\small Performance of DistPrivacy under different configurations.}
	\label{op}
	\vspace{-0.5cm}
\end{figure}

Figure \ref{nbl} illustrates the total latency incurred by the system for different privacy levels, when varying the number of participants. We note that the capacities of devices are uniformly distributed. When the number of devices is small, the total latency to compute all the requests is lower when the privacy level (SSIM) is equal to 0.8. This can be explained by the fact that only the shallow layers should be distributed on multiple devices (first 2 layers of  MNIST, first 8 layers of CIFAR and CAR, and first 4 layers of CELEBA). Furthermore, the privacy level can be achieved by dividing the resultant feature maps of the  shallow layers into only 2 devices. In this way, the total latency to offload the intermediate features is minimized and the shared data between devices is reduced as depicted in Figure \ref{nbd}. In case the tolerated privacy level is equal to 0.6 or 0.4, most of the layers should be distributed into a higher number of devices. Hence, low number of participants is not enough to satisfy security and resource availability. The percentages in Figure \ref{nbl} depict the proportion of requests that could not be processed due to the shortage of resources. In such a scenario, the system keeps trying to re-compile the rejected inferences, which justifies the high latencies and the low data load. When more devices participate in the distributed system, more resources are involved and tight privacy levels (SSIM= 0.6 or 0.4) can respect the security constraint. In this case, distributing the segments help to further parallelize tasks and reduce latencies compared to the 0.8 level. Meanwhile, when more participants contribute to the inference, the offloaded data increases (see Figure \ref{nbd}), incuring higher costs of transmissions and battery. To summarize, a trade-off between security and shared data/cost should be established as well as between the security level and latency when the number of devices is low.

Figures \ref{disl} and \ref{disd} present the total latency and shared data of the system, when varying the distribution of units' capacities. The results indicate that a pervasive system with limited-resources participants (70\% small devices (STM32H7) - 30\% powerful devices (RPi3)) is not adequate for distributed inference. When 50\% of devices or more are powerful, the system performs very well, as depicted in Figure \ref{disl}. Also, following the latter configuration, all requests are served, which is not the case of the 70\%-30\% system that rejects 29\% of the classification requests, when SSIM=0.4. Also, when SSIM = 0.8, much higher latency is incurred, as less tasks are parallelized and higher number of feature maps are computed in low-performance devices.

Figures \ref{typel} and \ref{typed} depict the total latency and the total shared data, when generating a set of requests comprising only one type of CNN. When all requests are classified by LeNet (7 layers and 28x28 images), the latency of the system and data load are very low as the network structure is only composed of 8 feature maps and involves 2 participants to ensure the privacy requirements. For a larger network (CIFAR CNN), the latency is still low compared to the time needed to process VGG16 and VGG19 requests, which is justified by the limited number of layers (12 layers), the small size of images (32x32), and the number of filters that does not exceed 128. VGG16 and VGG19 are well known for their performance on image classification thanks to their deep networks. However, their memory occupation and computation requirements restrain them from being processed on limited-resources devices. By distributing such models, high computations could be parallelized to achieve small decision-taking latency and the large memory occupation could be shared between participants. Collaborating to compute image classification, contributes not only to enhance the privacy of sensitive data and prevent it from inversion but also to reduce the latency, and pararellelize the CNN tasks, which is not the case of adding noise or encryption that require additional delays to achieve security. Note that low privacy level (SSIM=0.8) presents lower latency for VGG, because very high data sharing is required for tight privacies (SSIM=0.6 or 0.4) and some requests are rejected and re-sent to be classified when resources are available. Finally, our system is general enough to be implemented in multiple operating  units and different CNN networks, deployed in the same IoT system. 
\vspace{-0.1cm}
\section{Conclusion }\label{conclusion}
In this paper, we explored the feasibility of recovering private  data of distributed CNN using black-box attack. We discovered that distributing the feature maps into more participants can strengthen the privacy of the origin image. Therefore, we re-designed the deep learning solutions, which require high computational and memory demands, to match the constraints characterizing
IoT units, aiming at minimizing the black-box risks. This method has been formulated as an optimization problem, where classification latency is reduced. Next, due
to the problem complexity, we proposed an online solution that is adequate for real-time scenarios. Our simulation unveiled different parameters that should be present to achieve the privacy of distributed CNNs, including the number of devices, their capacities, and the deployed networks. Future works will encompass a comparison with existing distributed CNN systems and security countermeasures
\bibliographystyle{IEEEtran}
\bibliography{IEEEabrv,IEEEexample}

\end{document}